\documentclass[aps,twocolumn,preprintnumbers]{revtex4}

\usepackage{graphicx}  
\usepackage{subfigure}
\usepackage{multirow}
\usepackage{tikz}

\usepackage{fancyhdr}
\usepackage{longtable}
\usepackage{parskip}
\usepackage[T1]{fontenc}
\usepackage{dcolumn}   

\usepackage{bm}        
\usepackage{amsfonts}  
\usepackage{amsmath}   
\usepackage{amssymb}   

\usepackage{color}
\usepackage[dvipsnames]{xcolor}
\usepackage{xurl}
\usepackage[hidelinks]{hyperref}

\DeclareRobustCommand{\quantumlegend}{%
  \tikz[baseline=-0.5ex]\draw[dash dot dot, thick] (0,0) -- (1,0);%
}
\DeclareRobustCommand{\classicallegend}{%
  \tikz[baseline=-0.5ex]\draw[dash pattern=on 6pt off 2pt, thick] (0,0)--(0.8,0);%
}
\DeclareRobustCommand{\replegend}{%
  \tikz[baseline=-0.5ex]\draw[dash pattern=on 4pt off 2pt, line width=1.2pt] (0,0) -- (1.1,0);%
}

\newcommand{\bra}[1]{\left\langle #1 \right|}

\newcommand{\ket}[1]{\left| #1 \right\rangle}

\def\CC{\mathbb{C}}
\newtheorem{example}{Example}

\begin{document}
\title{A hybrid variational quantum circuit approach for stabilizer states classifiers}

\author{Hamna Aslam$^{\dagger}$, Fr\'ed\'eric Holweck$^{\dagger,\ddagger}$}

\affiliation {\it ${}^\dagger$ Laboratoire Interdisciplinaire Carnot de Bourgogne,  UMR 6303 CNRS, University of Technology of Belfort-Montb\'eliard, 90010 Belfort Cedex, France}
\affiliation{\it ${}^\ddagger$ Mathematics and Statistics Department, Auburn University, Auburn, AL, USA}

\date{\today}

\begin{abstract}  
Entanglement classification of pure multipartite quantum states is a challenging problem in quantum information theory that can be mathematically stated as orbit classification for some given group action on the ambient Hilbert space. The group action depends on the grained classification one expects, the finer-grained one being the classification up to local unitary transformation (LU). In this article, we show how a variational quantum circuit approach can be used to learn entanglement orbits, and we apply our findings to build a classifier for four-qubit states. 
\end{abstract}


\maketitle

\section{Introduction}

Quantum entanglement has gained much attention in recent years due to its role as a fundamental resource in many quantum information applications including quantum simulation \cite{dechiara}, quantum error correction \cite{calderbank}, quantum teleportation \cite{bennet}, and quantum communication \cite{ekert}. Quantum computation protocols that demonstrate a quantum speed-up in problem solving also rely heavily on entanglement \cite{shor, grover}. The characterization and classification of entangled quantum states is thus of significant importance and has been widely studied \cite{eltschka, wang1, rangamani, acin}. One of the most common methods of entanglement classification is based on equivalence relations of local operations, which often include Local Unitary (LU), Local Operations and Classical Communications (LOCC), and Stochastic Local Operations and Classical Communications (SLOCC) \cite{dur, verstraete}. When restricting to graph states, the group of Local Clifford operations (LC), a subgroup of LU, is relevant. Under this framework, all states that are related by the action of a group (e.g. LU, LOCC, SLOCC, LC) are said to belong to the same orbit or the same equivalence class. Being a non-local property, the action of local operations leaves the intrinsic nature of entanglement unchanged and states belonging to the same orbit have similar entanglement properties.

Classifying an unknown quantum state based on its orbit often requires a complete knowledge of the state. Even classical machine learning methods, that have been successfully employed for this purpose \cite{jaffali, varela, asif}, require some information of the quantum state that can be used as input to the model, which can be very challenging due to the closed nature of quantum systems. One way to obtain this information is through quantum state tomography \cite{paris}. This method can be very noisy and resource intensive since it requires measurements whose numbers scale exponentially with the number of subsystems. Another method called quantum state discrimination, which uses Positive Operator-Valued Measurements (POVM), can also be used to gain information about a state \cite{bae}. Unfortunately, finding the optimal POVMs can be a very complicated and resource intensive task. 

In light of these difficulties, a variational quantum circuit approach to entanglement classification appears to be a natural solution. When using a Variational Quantum Classifier (VQC), the quantum circuit operations can be applied directly on the quantum system to be classified, thereby eliminating the need for information extraction through resource intensive measurements. The natural Hilbert space in which these quantum circuits operate also allows a polynomial scaling of the resources with the number of subsystems. Furthermore, since VQCs outsource the computationally intensive optimization to a classical computer, they can be readily executed on Noisy Intermediate-Scale Quantum (NISQ) era hardware. 

Extensive prior work has been done on variational quantum classification of entangled states \cite{scala, qiu, schatzki, grant}. However, most of it uses classification methods based on entanglement witnesses or measures. While there have been attempts at classifying entanglement based on local operation orbits \cite{wang2}, they have been restricted in a manner that does not explore the full equivalence class and reduces the problem to a linear classification problem.

The rest of the paper is arranged as follows with Section \ref{background} giving some basic background on VQCs and entanglement classification of pure states. Section \ref{learning_orbits} discusses the non-linearity of orbits and the manner in which a hybrid VQC approach can learn these non-linear orbits much better than a simple VQC. Section \ref{implementation} presents the implementation details of the hybrid VQC and Section \ref{applications} demonstrates our approach by applying this hybrid VQC to learn various classification schemes of four-qubit graph states including the LC and LU orbit identification. Finally, Section \ref{conclusion} presents the conclusions.
\section{Background}
\label{background}
We review in this section some basic notions on variational quantum circuits and entanglement classification of pure states
\subsection{VQC}
The classification of entangled states using a Variational Quantum Classifier (VQC) can be categorized as a supervised quantum machine learning task \cite{mitarai, schatzki}. The first essential component of such tasks is a labeled dataset of the form $\{\boldsymbol{x}_i,y_i\}$ with $\boldsymbol{x}_i$ being the feature vector of the input quantum state $\ket{\psi_i}$, and $y_i$ being the corresponding label. Here, the feature vector $\boldsymbol{x}_i$ is the $d$-dimensional amplitude vector of the quantum state $\ket{\psi_i}\in \mathcal{S}\subseteq \mathcal{H}$, such that $\ket{\psi_i}$ is an $n$-qubit pure quantum state with $d=2^n$. Since we only focus on binary classification, the labels considered are $y_i\in\{-1,1\}$. The VQC acts as a parametrized function $f(\boldsymbol{x}_i;\boldsymbol{\theta})$ which is trained to predict the labels by varying $\boldsymbol{\theta}$, with the goal often being the generalization of this function to unseen datasets.

\begin{figure}[h!]
  \centering
  \includegraphics[width=0.45\textwidth]{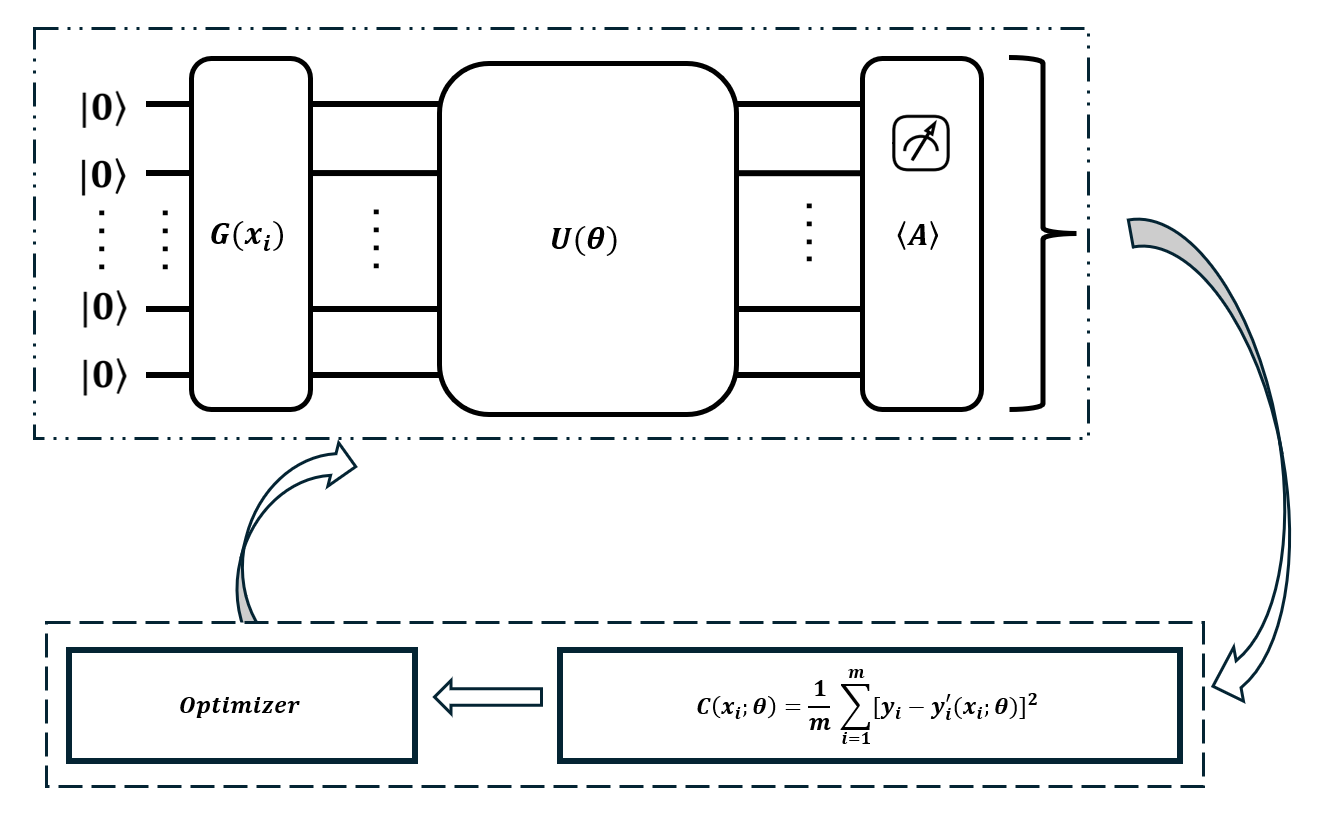}
  
\caption{ Schematic diagram of a general $n$-qubit simple Variational Quantum Classifier. The pattern  
  \quantumlegend indicates execution on a quantum device and \classicallegend
indicates execution on a classical device.}
  \label{fig:vqc_circuit}
\end{figure}

A schematic representation of a general VQC is shown in FIG. \ref{fig:vqc_circuit}, where the first block of operations $G(\boldsymbol{x}_i)$ represents the encoding of the feature vector $\boldsymbol{x}_i$ into the quantum circuit. Following that, repeated layers of some combination of parametrized rotation gates and entangling gates, represented by $U(\boldsymbol{\theta})$ are implemented. Finally, a measurement over some observable $A$ is conducted to extract the model's predicted label $y'_i(\boldsymbol{x_i};\boldsymbol{\theta})$ from the quantum circuit

\begin{equation}
\langle A\rangle=\bra{\psi_o}G^{\dagger}(\boldsymbol{x}_i)U^{\dagger}(\boldsymbol{\theta})\,A\,U(\boldsymbol{\theta})G(\boldsymbol{x}_i)\ket{\psi_o}=y'_i(\boldsymbol{x_i};\boldsymbol{\theta})
\end{equation}
\\
where $\ket{\psi_o}=\ket{00..0}$ is the initial state of the circuit. To quantify the success of the model's prediction, a cost function is defined as

\begin{equation}
    C(\boldsymbol{x_i};\boldsymbol{\theta})=\frac{1}{m}\sum_{i=1}^{m}(y_i-y'_i(\boldsymbol{x_i};\boldsymbol{\theta}))^2
\end{equation}
\\
with $m$ being the total number of samples in the dataset. This makes the task of classification into an iterative optimization task such that

\begin{equation}
\boldsymbol{\theta}_{opt}=\text{arg\,min}_{\boldsymbol{\theta}}\,\,C(\boldsymbol{x_i};\boldsymbol{\theta})
\end{equation}
\\
Within the VQC, both the cost function calculation and optimization are executed on a classical computer.

\subsection{Entanglement classification}
One knows since the early 2000s that for $n \geq 3$, pure quantum states can exhibit entanglement in non-equivalent ways \cite{dur}. \emph{Non-equivalent} means that two states cannot be transformed into each other by local operations. Conversely, \emph{equivalent} states are those that can be interconverted by local operations, and are therefore regarded as belonging to the same entanglement class. All states $\ket{\psi'}$ equivalent to a given state $\ket{\psi}$ under a group action $G$ are said to belong to the $G$-orbit of $\ket{\psi}$. The type of local operations considered depends on the desired granularity of the classification. Let us first introduce two standard schemes of classifications:

\begin{enumerate}
    \item \textbf{Local Unitary (LU)}: LU$=\text{SU}_2(\CC)^{\times n}$.  
    Here the group LU acts on $\mathcal{H}_n$ as the product of local unitaries. We say $\ket{\psi} \equiv_{\text{LU}} \ket{\psi'}$ if and only if there exist $A_i \in \text{SU}_2(\CC)$, $i=1,\dots,n$, such that  
    \[
    \ket{\psi} = A_1 \otimes \dots \otimes A_n \ket{\psi'}.
    \]  
    The LU classification is a fine grained one. Note that for pure states, LU equivalence coincides with LOCC (Local Operations with Classical Communication \cite{dur}) equivalence.

    \item \textbf{Stochastic Local Operations and Classical Communication (SLOCC)}: SLOCC$=\text{SL}_2(\CC)^{\times n}$.  
    Here the restriction to unitaries is relaxed: local operators are only required to be invertible. Thus,  $\ket{\psi} \equiv_{\text{SLOCC}} \ket{\psi'}  \Leftrightarrow \exists A_i \in \text{SL}_2(\CC), i=1,\dots,n, \text{such that }$
    \[
     \ket{\psi}=A_1\otimes \dots \otimes A_n \ket{\psi'}.
    \]  
\end{enumerate}

\begin{example}
Up to LU equivalence, any two-qubit pure state $\ket{\psi}\in \mathcal{H}_2$ can be brought to its Schmidt form  
\[
\ket{\psi} \equiv_{\text{LU}} \sqrt{\lambda_0}\ket{00} + \sqrt{\lambda_1}\ket{11},
\]  
with $\lambda_0,\lambda_1 \in \mathbb{R}_+$ and $\lambda_0+\lambda_1=1$. This yields a continuum of LU classes parametrized by $\lambda_1 \in [0,1/2]$: the case $\lambda_1=0$ corresponds to the class of separable states, while $\lambda_1=1/2$ corresponds to the maximally entangled class represented by the well-known EPR state \cite{nielsen},  
\[
\ket{EPR} = \frac{1}{\sqrt{2}}(\ket{00}+\ket{11}) .
\]  
In contrast, up to SLOCC equivalence, there are only two orbits: the orbit of separable states (all states SLOCC-equivalent to $\ket{00}$) and the orbit of entangled states (all states equivalent to $\ket{EPR}$).
\end{example}

\begin{example}
For $n=3$, LU equivalence allows one to reduce a three-qubit state to the form  
\[
\ket{\psi} = \lambda_0\ket{000} + \lambda_1 e^{i\phi}\ket{100} + \lambda_2\ket{101} + \lambda_3\ket{110} + \lambda_4\ket{111},
\]  
with $\lambda_i \geq 0$, $\phi \in [0,2\pi]$, and $\sum \lambda_i^2 = 1$, leading again to infinitely many LU orbits \cite{acin}.  

For the SLOCC classification \cite{dur}, the three-qubit Hilbert space decomposes into six classes:  
\begin{itemize}
    \item the orbit of separable states ($\text{SLOCC}.\ket{000}$),  
    \item three biseparable orbits ($\text{SLOCC}.\ket{0}\otimes \ket{EPR}$, $\text{SLOCC}.\tfrac{1}{\sqrt{2}}(\ket{000}+\ket{101})$, $\text{SLOCC}.\ket{EPR}\otimes \ket{0}$),  
    \item the $W$ class \big($\text{SLOCC}.\tfrac{1}{\sqrt{3}}(\ket{100}+\ket{010}+\ket{001})$\big),  
    \item and the $GHZ$ class \big($\text{SLOCC}.\tfrac{1}{\sqrt{2}}(\ket{000}+\ket{111})$\big).
\end{itemize}
\end{example}

For $n \geq 4$, both LU and SLOCC classifications contain infinitely many orbits \cite{verstraete}. The SLOCC classification of pure states remains tractable for $n=4$ \cite{holweck1,holweck2}, but becomes intractable in full generality for $n \geq 5$ \cite{luque}.

\section{Learning orbits}
\label{learning_orbits}
\subsection{Non-linearity}

Consider a group of invertible linear transformation $G$ and a state $\ket{\psi}\in \mathcal{H}$. The $G$-orbit of $\ket{\psi}$ is the set of all states equivalent to $\ket{\psi}$, we denote this set by $G\cdot\ket{\psi}$. Consider now the set of all polynomials on $\mathcal{H}$ that vanish on $G\cdot\ket{\psi}$ and denote by $\mathcal{I}_{\ket{\psi}}$ this set. Then, the orbit closure of the orbit $G\cdot\ket{\psi}$, which is denoted by $\overline{G\cdot\ket{\psi}}$, is the set of elements of $\mathcal{H}$ that annihilate all polynomials in $\mathcal{I}_{\ket{\psi}}$. Clearly $G\cdot\ket{\psi}\subset \overline{G\cdot\ket{\psi}}$. Such set, the vanishing of a family of polynomials, is also known as an algebraic variety.
For G=LU and G=SLOCC, such polynomials are typically non-linear and the vanishing of these polynomials defines the decision boundaries between the equivalence classes or orbits \cite{jaffali, turner}.
For example, in the simplest case of a pure two-qubit quantum states, it is well-known that the separation between separable and entangled states is given by the vanishing of the $2\times 2$ determinant, i.e. $\ket{\psi}=\sum_{i,j\in \{0,1\}} a_{ij}\ket{ij}$ is separable if and only if $a_{00}a_{11}-a_{01}a_{10}=0$. In the three-qubit case, the union of the SLOCC orbits other than the $\ket{GHZ}$-one is given by the vanishing of a degree 4 invariant, known in the mathematics literature as Cayley hyperdeterminant \cite{holweck3, djokovic}. The degree of this polynomial increases exponentially with the number of qubits \cite{gkz}.

Since such separations between equivalence classes or orbits are defined by the vanishing of non-linear polynomials, the boundaries of classification that the VQC needs to recognize are correspondingly non-linear. This is especially difficult for simple VQCs since all the quantum gates applied within the quantum circuit are linear unitary transformations and non-linearity is only provided towards the end by measuring qubits. A simple VQC thus struggles to approximate the highly non-linear decision boundaries that are required by our classification task.

This difficulty is demonstrated in FIG. \ref{fig:synthetic_dataset} with the use of a simple synthetic linearly non-separable two-dimensional dataset. The dataset was created by randomly generating two-dimensional data points $(x,y)$ with $x,y \in [-1,1]$. The data points were then labeled to be a blue cross or a red point such that the boundaries of classification were non-linear, as shown in in FIG. \ref{fig:synthetic_dataset} (a). The classification performance of a simple one-qubit VQC, with the data samples amplitude-encoded into it, is shown in FIG. \ref{fig:synthetic_dataset} (b). It can be observed that this VQC only manages to draw linear classification boundaries, misclassifying many red points as blue crosses.

\subsection{A hybrid VQC}

\begin{figure}[h!]
  \centering
  \subfigure[]{
    \includegraphics[width=0.22\textwidth]{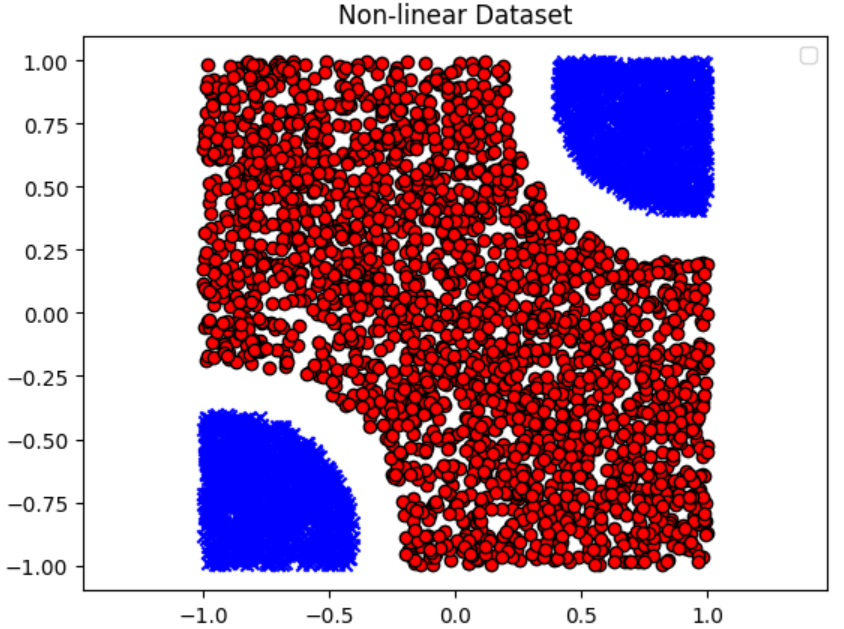}
    \label{fig:nonlinear_dataset}
  }
  \subfigure[]{
    \includegraphics[width=0.22\textwidth]{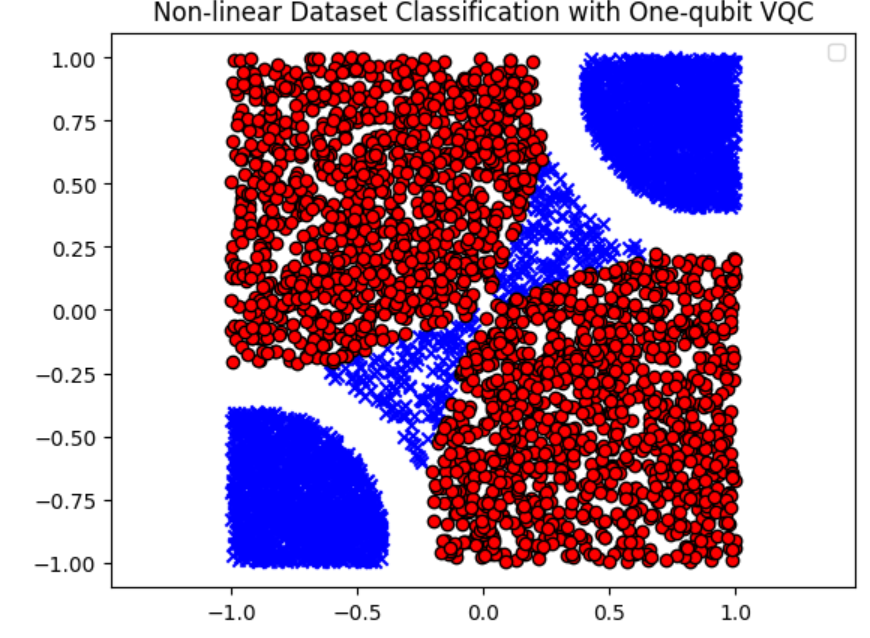}
    \label{fig:nonlinear_one_qubit}
  }
  \subfigure[]{
    \includegraphics[width=0.22\textwidth]{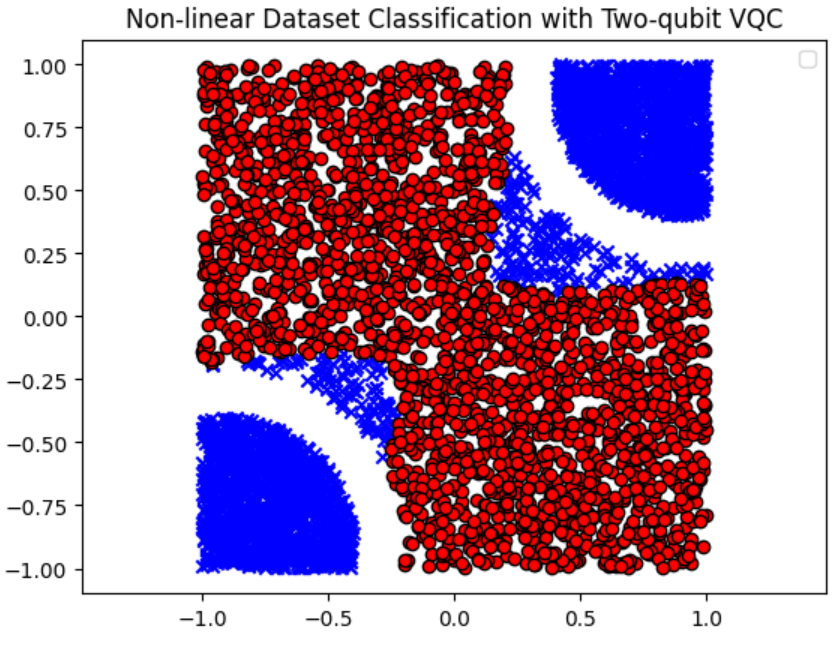}
    \label{fig:nonlinear_two_qubits}
  }
  \subfigure[]{
    \includegraphics[width=0.22\textwidth]{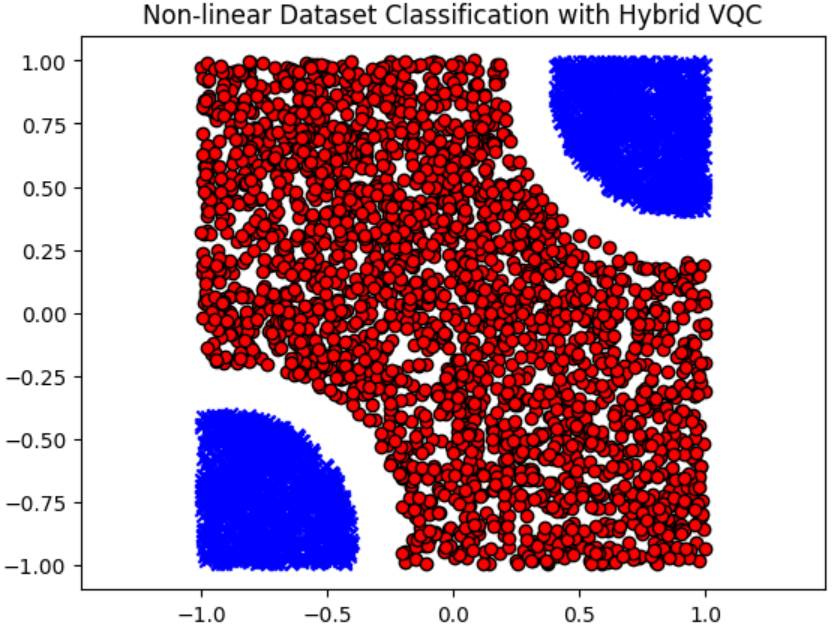}
    \label{fig:nonlinear_hybrid}
  }
  \caption{(a) shows the original classification of the dataset. (b) shows the results when a single-qubit VQC is trained to classify this dataset. It can be observed that this VQC can only draw linear boundaries in its attempt to minimize the cost function. (c) shows the results when a two-qubit VQC is trained to classify the dataset. The two features of the dataset were encoded into the amplitudes \cite{rath} of the qubits with the other two amplitudes treated as free hyperparameters and set to 0 and 0.25. However, it is obvious that even with these additional degrees freedom, this VQC cannot fully grasp the non-linearity required from the problem. Finally, (d) shows the performance of a two-qubit hybrid VQC that conducts classical post-processing using classical NN layers. The classification accuracy is $100\%$ for this case.}
  \label{fig:synthetic_dataset}
\end{figure}

There are many ways to deal with this non-linearity issue in VQCs, the most notable of which include non-linear feature maps \cite{mitarai} and exploitation of the tensor product structure of the quantum circuit \cite{schuld}. Unfortunately, non-linear feature maps often require full knowledge of the input feature vector, which is very resource-intensive as mentioned before, and exploitation of tensor product structure requires significantly scaling up the number of qubits. 
Another way to improve the non-linear performance of the VQC is by increasing its expressivity via an increase in the quantum circuit's width or depth \cite{mangini}. FIG. \ref{fig:synthetic_dataset} (c) shows the employment of this method where the number of qubits is increased from one to two such that the circuit can have four features amplitude-encoded into it. The first two features are the data sample features and the other two features are treated as free parameters to improve expressivity. It can be observed that while this introduces some non-linearity into the boundaries, it is still not enough to correctly classify the dataset at hand. Thus, for accurate classification of highly non-linear boundaries, this method also requires significantly scaling up either the number of qubits or the number of implemented gates, neither of which is ideal in the NISQ era.

A method that does not have any of these drawbacks is the use of classical neural network (NN) layers for post-processing. To conduct this post-processing, a separate measurement can be performed on each of total $n$ qubits and the resulting $n$-dimensional output vector $\boldsymbol{y'_i}(\boldsymbol{x_i};\boldsymbol{\theta})$ can then be passed on to classical NN layers as input. The number of layers and the number of neurons in each of these layers can be varied. If there are $h$ neurons in the first layer and $W\in\mathbb{R}^{h\,\mathsf{x}\,n}$, $\boldsymbol{b}\in\mathbb{R}^{h}$ are trainable parameters, then the action of the first layer involves the implementation of a non-linear activation function $\sigma$ as 

\begin{equation}
\boldsymbol{y'_i(\theta})\xrightarrow{\;\;layer\;\;}\sigma(W.\boldsymbol{y'_i(\theta})+\boldsymbol{b})
\end{equation}
\\
Here, $\sigma$ acts element-wise on the entries such that $\sigma(\boldsymbol{z})=[\sigma(z_1), \sigma(z_2),...]$. The non-linearity provided by this activation function's implementation on the data during post-processing can significantly improve the action of the classifier on linearly non-separable datasets. This can be seen in FIG. \ref{fig:synthetic_dataset} (d), where a hybrid VQC was employed with post-processing via a single classical neuron and the classification accuracy was $100\%$.

Once this post-processing is done, the output of the final classical NN layer: ${y^{C}_i(\boldsymbol{x_i};W,\boldsymbol{b,\theta}})$, can be provided to the cost function and the optimizer can update the trainable parameters in the quantum circuit as well as the classical NN layers. 

This type of post-processing not only improves the ability of the VQC to classify linearly non-separable datasets, it also increases the number of trainable parameters significantly without increasing the width or depth of the quantum circuit. Thus, this architecture is especially well-suited for the complex task of classifying the local operations' non-linear orbits. 

This suitability can be demonstrated with a very simple example where states are classified by the hybrid VQC into either the LU orbit of the three-qubit GHZ state or the LU orbit of one of the other three-qubit states mentioned in TABLE \ref{tab:GHZ_results}. The same table shows the results of this classification, revealing that the hybrid VQC can indeed identify the states and thus learn the non-linear orbits with very high accuracy. However, since the classifier was trained to recognize only two LU orbits, instead of identifying one LU orbit from the full state space, this problem is relatively simple. A more thorough and complicated approach can thus be taken by training the hybrid VQC to learn and classify an LU orbit from the entire Hilbert space. The results for this type of classification are shown in TABLE \ref{tab:3_qubit_LU_results}, which show that the classifier also does a good job of identifying the non-linear orbits even when presented with random states from the full state space.
\begin{table}[!ht]
\begin{tabular}{|c|c|c|c|}
\hline
State Name  & Training Accuracy  & Test Accuracy \\
\hline
Fully Separable & $99\%$ & $98\%$\\
\hline 
Biseparable (AB-C)  & $98\%$          & $97\%$ \\
\hline 
Biseparable (A-BC)  &$98\%$ & $97\%$ \\
\hline 
Biseparable (B-AC)  & $99\%$ & $99\%$ \\
\hline 
W  &$100\%$  & $100\%$ \\
\hline
\end{tabular}
\caption{Training and test accuracy results for the classification of LU orbit of three-qubit GHZ state $(\frac{1}{\sqrt{2}}(\ket{000}+\ket{111}))$ against the LU orbits of some other well known three-qubit states.}
\label{tab:GHZ_results}
\end{table}

\begin{table}[!ht]
\begin{tabular}{|c|c|c|c|}
\hline
State Name  & Training Accuracy  & Test Accuracy \\
\hline
Fully Separable & $90\%$ & $89\%$\\
\hline 
Biseparable (AB-C)  & $98\%$          & $97\%$ \\
\hline 
Biseparable (A-BC)  &$100\%$ & $99\%$ \\
\hline 
Biseparable (B-AC)  & $97\%$ & $95\%$ \\
\hline 
W  & $84\%$ & $83\%$ \\
\hline 
GHZ  &$97\%$  & $97\%$ \\
\hline
\end{tabular}
\caption{Training and test accuracy results for the classification of LU orbits of various three qubit states against the full 3 qubit Hilbert space.}
\label{tab:3_qubit_LU_results}
\end{table}

\section{Implementation}
\label{implementation}
\begin{figure}[h!]
  \centering
  \includegraphics[width=0.5\textwidth]{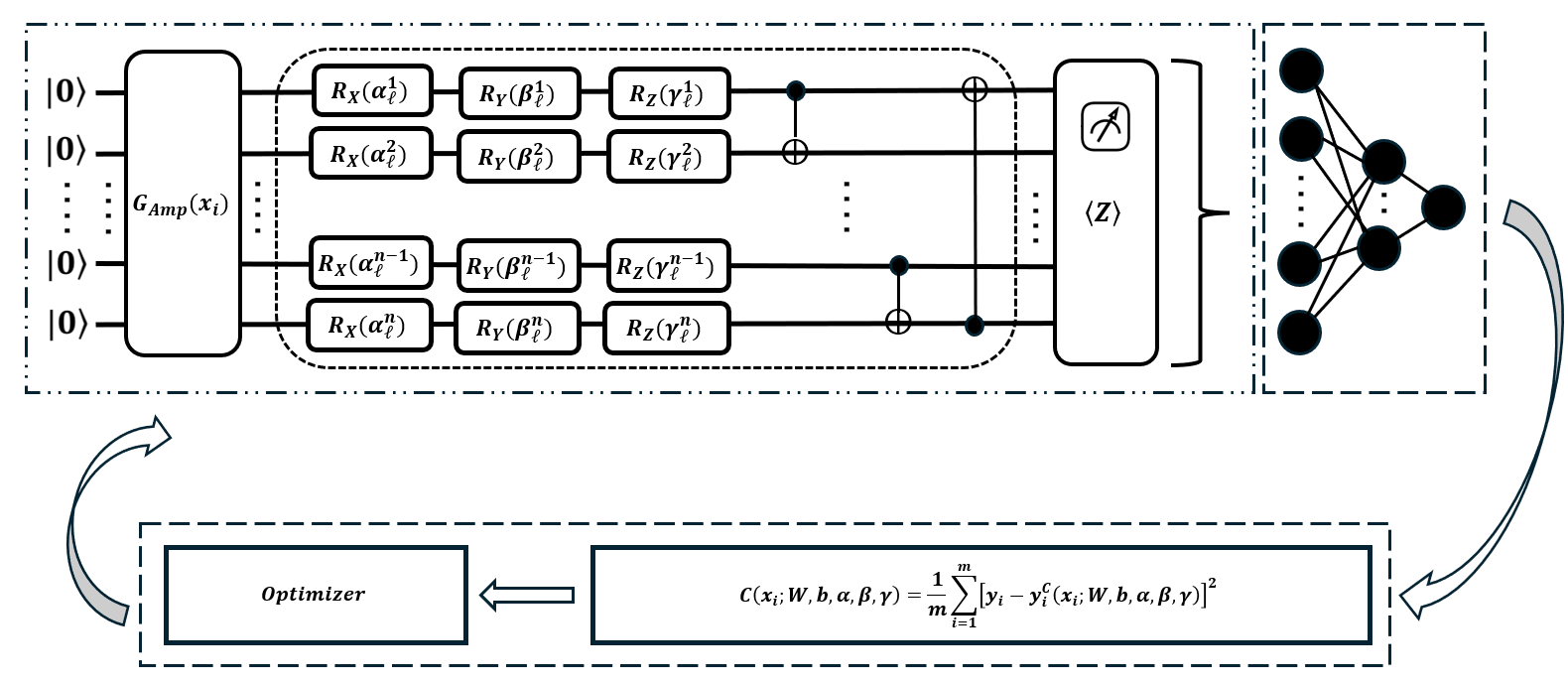}
  \caption{Schematic $n$-qubit diagram of the specific hybrid VQC structure employed in this work. Here, $\boldsymbol{\theta}=\{\boldsymbol{\alpha,\beta,\gamma}\}$ form the set of trainable parameters of the quantum circuit with their upper and lower indices indicating the qubit number and layer number respectively. $G_{AMP}$ indicates amplitude encoding \cite{rath} and $R_X,R_Y,R_Z$ are single qubit rotation gates round $X,Y,Z$ axes respectively. The diagram only shows one hidden layer of the classical NN for simplicity. In practice, the number of hidden layers and the number of neurons in the hidden layers were varied for each classification task. The pattern \quantumlegend indicates execution on a quantum device, \replegend indicates repeated layer of the quantum circuit, and \classicallegend indicates execution on a classical device.}
  \label{fig:hybrid_vqc_circuit}
\end{figure}

A schematic diagram of the specific hybrid VQC structure employed in this work, along with its classical post-processing layers is shown in FIG. \ref{fig:hybrid_vqc_circuit} for the general case of $n$-qubits.

In practice, the hybrid VQC was implemented in PennyLane using their \texttt{TorchLayer} class that allows the variational circuit to be treated as a PyTorch layer, which can then be used in combination with classical NN layers to create hybrid models within PyTorch's \texttt{Sequential} or \texttt{Module} classes. 

First, the quantum model, consisting of the full quantum circuit, including its variational parameters, was defined in PennyLane. The parameters and the model were then wrapped into a Torch layer using the \texttt{TorchLayer} class. The layers of the classical NN were then generated using the \texttt{Linear} class in PyTorch with the number of hidden layers and the number of neurons in each hidden layer varying depending upon the classification task. These layers were combined with the activation functions as well as the quantum model layer through the \texttt{Sequential} class to define the final hybrid VQC model. The number of layers in the quantum circuit was kept $4$ but the number of hidden layers in the classical NN varied from $1$ to $2$, with the number of neurons in these hidden layers ranging from $50$ to $500$, all depending on the task and found through trial and error.

Since the outputs of the variational circuit were expectation values ranging from $-1$ to $1$, the $tanh$ activation function was used in the classical layers. Furthermore, considering the fact that the circuit's output was expectation values and not probabilities, the Mean Squared Error cost function was used instead of the Binary Cross Entropy cost function (which requires probabilistic inputs). The optimizer used was Adam and the learning rate was varied between $0.001$ and $0.002$ for each learning task.

After defining the model, the training loop was initiated. Batches of the dataset, ranging in size from 10 to 20 samples, were loaded into the model using PyTorch's \texttt{DataLoader} class and the parameters of the quantum circuit as well as the classical NN layers were optimized to minimize the cost function in each training step using the \texttt{.step()} method. The total number of training samples varied for each classification task but often ranged from $3000$ to $32400$ such that each classification task began with $3000$ training samples, and this number was increased until the final training and test accuracies showed similar results and until further increasing the number didn't improve the average training accuracy. Note that the number of training and test samples was always kept equal.

Once trained, the model was asked to predict values for all training samples and the training accuracy was evaluated by defining a custom accuracy function within which all predicted values less than $0$ were assigned class label $-1$ and all predicted values greater than $0$ were assigned class label $+1$. The accuracy value was then given by squaring the sum of the predicted class label and actual class label and normalizing the result. The same method was used to evaluate test accuracy on unseen test samples. The accuracy values reported here are averaged over three independent runs of the VQC with all standard deviations $\leq3\%$ and then rounded off to the nearest percent.
\section{Applications}
\label{applications}
To illustrate our approach better, we apply it to the study of orbits of four-qubit graph states under different classification schemes of varying complexity.  
First, we consider the set of all four-qubit graph states and show how our hybrid VQC can be trained to distinguish their entanglement classes (\ref{sec:graph}).  
Next, instead of the orbits of graph states themselves, we examine the orbits of graph states under Local Clifford ($LC$) operations. For four-qubits \footnote{False for $n=27$ qubits but true for $n\leq 19$ \cite{perdrix}}, the orbit $LC \cdot \ket{\psi_G}$ of a given graph state $\ket{\psi_G}$ corresponds to the set of all stabilizer states $LU$-equivalent to $\ket{\psi_G}$ (\ref{sec:stabilizer}).  
Furthermore, we consider the full $LU$-orbits of the six representatives of the four-qubit graph state classification and demonstrate that our trained circuit successfully distinguishes them for each other but also from any generic states (\ref{sec:lu-four-qubit}). Finally, we scale up our method to larger systems of 5 and 6 qubits, and classify the LU equivalence class of the their star graph states against the rest of their LU equivalence classes, limited to connected graphs. We show that our method works well for larger number of qubits but an increased number of samples and number of hidden neurons in the post-processing NN are required (\ref{sec:star_graph}).

\subsection{The four-qubit graph states classification}\label{sec:graph}
Graph states are multipartite quantum states described by a graph.  
Consider a graph $G=(V,E)$ with $|V|$ vertices labeled by integers $i\in\{0,\dots, |V|-1\}$ and $|E|$ edges. The graph state $\ket{\psi_G}$ associated with $G$ is a $|V|$-qubit quantum state defined by
\[
\ket{\psi_G}=\prod_{e\in E} CZ_{e}\cdot \ket{+}^{\otimes |V|},
\]
where $CZ_{e}$ is a controlled-$Z$ gate acting on qubits $i$ and $j$ when $e$ is the edge connecting vertices $i$ and $j$.  

Graph states play an important role in quantum information theory. They have applications in quantum error correction \cite{hein}, quantum secret sharing \cite{markham}, and measurement-based quantum computation.  

\begin{figure}[!ht]
\includegraphics[width=0.4\textwidth]{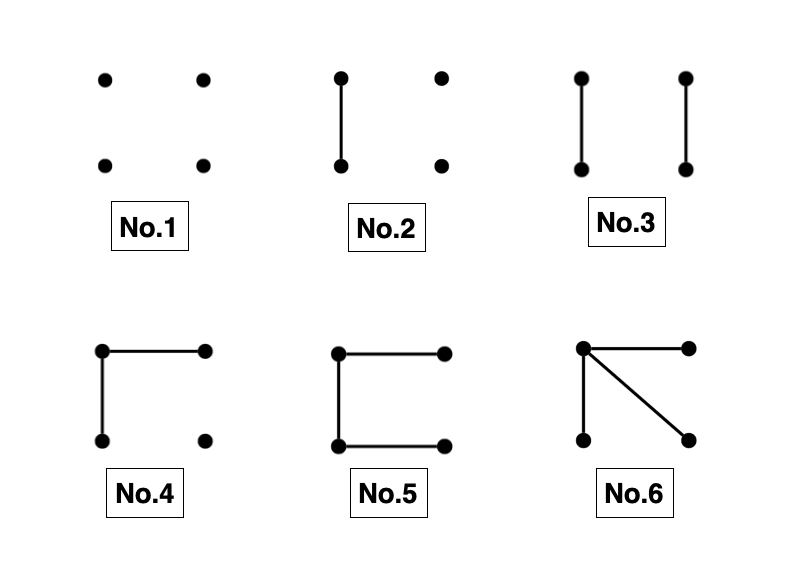}
\caption{Classification of four-qubit graph states. Note that the star graph corresponds to the four-qubit GHZ state.}
\label{fig:graph_states_1}
\end{figure}

Up to local transformations, a classification of graph states is known for up to $n=12$ qubits \cite{cabello}.  
For four qubits, there are $2^6=64$ different graph states, which fall into six distinct equivalence classes under local transformations (FIG. \ref{fig:graph_states_1}).

The canonical form of those graph states is $\ket{\psi_G}=\sum_{x_1,x_2,x_3,x_4\in \{0,1\}} (-1)^{f_G(x_1,x_2,x_3,x_4)}\ket{x_1x_2x_3x_4}$ where the function $f_G$ is a boolean function with monomial $x_ix_j$ appearing for each $CZ_{ij}$. In the following table we provide examples of four-qubit states $LU$ equivalent to each representative of the class as well as the number of graphs equivalent to it (up to permutation and local operations).

\begin{table}[!ht]
\begin{tabular}{|c|c|c|c|}
\hline
class & $f_G(x_1,x_2,x_3,x_4)$  & \# of $G$-states  &LU-equi \\
 N°     &        &   in the class & four-qubit state\\
\hline
$1$ & $0$ & $1$& $\ket{0000}$\\
\hline 
$2$   & $x_1x_2$          & $6$&  $\ket{EPR}\otimes \ket{00}$  \\
\hline 
$3$ &$x_1x_2+x_3x_4$ & $3$ &$\ket{EPR}\otimes \ket{EPR}$ \\
\hline 
$4$ & $x_1(x_2+x_3)$ & $16$ & $\ket{GHZ}\otimes \ket{0}$\\
\hline 
$5$ & $x_1x_2+x_2x_3+x_3x_4$  & $33$ & $\frac{1}{2}(\ket{0000}+\ket{0111}$\\
 & $+x_3x_4$  & & $+\ket{1011}+\ket{1100})$\\
\hline 
$6$ &$x_1x_2+x_1x_3+x_1x_4$  & $5$ & $\ket{GHZ_4}$\\
\hline
\end{tabular}
\caption{Class, example of representative (normal form) and number of equivalent graph states}
\label{tab:graph_state_classes}
\end{table}

We trained our hybrid VQC to identify these six different classes of four-qubit graph states. For the dataset, the amplitude vectors of all possible graph states were first generated. Then $50\%$ of the samples, labeled -1, were picked from the class to be identified and the other $50\%$, all labeled 1, were picked uniformly from the rest of the classes. Note that, due to the limited number of graph states, repetitions were allowed while picking the samples. 

The dataset was split evenly into training and test samples, with the results of the classification shown in TABLE \ref{tab:graph_states_reults}. It can be observed that our hybrid VQC does an excellent job of identifying each class with an accuracy close to $100\%$ in almost all cases.

\begin{table}[!ht]
\begin{tabular}{|c|c|c|c|}
\hline
Class N° & Training Accuracy  & Test Accuracy \\
\hline
$1$ & $100\%$ & $100\%$\\
\hline 
$2$   & $100\%$          & $100\%$ \\
\hline 
$3$ &$99\%$ & $99\%$ \\
\hline 
$4$ & $98\%$ & $98\%$ \\
\hline 
$5$ & $98\%$  & $98\%$ \\
\hline 
$6$ &$100\%$  & $100\%$ \\
\hline
\end{tabular}
\caption{Training and test accuracy results for identification of entanglement classes of four-qubit graph states.}
\label{tab:graph_states_reults}
\end{table}

\subsection{The four-qubit stabilizer states classification}\label{sec:stabilizer}

The $n$-qubit Clifford group $\mathcal{C}_n$ is the group of $n \times n$ unitary matrices that map the Pauli Group $\mathcal{P}_n$ to itself under conjugation \cite{zeng}. Two $n$-qubit states $\ket{\psi}$ and $\ket{\psi'}$ are Local Clifford (LC) equivalent if there exist operations $C_1,C_2,...\,,C_n$ within the Clifford group such that

\begin{equation}
    \ket{\psi'}=C_1\otimes C_2\,\otimes\,...\otimes\, C_n\ket{\psi}
    \label{eq:clifford_eq}
\end{equation}

Any stabilizer state is LC equivalent to some graph state \cite{nest}, hence, the full space of $n$-qubit stabilizer states can be reached by applying LC operations to the full set of $n$-qubit graph states. 

We trained the hybrid VQC to recognize the LC orbit of each of the four-qubit graph states represented in FIG. \ref{fig:graph_states_1}. This was done by first generating all 64 graph states and then applying random uniformly picked LC operations to generate samples of stabilizer states. The \texttt{random\_clifford} function provided by Qiskit (v0.37) was used to create these LC operations. As before, the dataset was kept balanced: $50\%$ of the sample stabilizer states were picked from the orbit to be learned and the other $50\%$ were picked uniformly from the rest of the 5 orbits. TABLE \ref{tab:stabilizer_states_reults} shows the results of the training and test accuracies for this case.

\begin{table}[!ht]
\begin{tabular}{|c|c|c|c|}
\hline
Class N° & Training Accuracy  & Test Accuracy \\
\hline
$1$ & $97\%$ & $96\%$ \\
\hline 
$2$   & $98\%$          & $98\%$ \\
\hline 
$3$ &$90\%$ & $90\%$ \\
\hline 
$4$ & $95\%$ & $95\%$ \\
\hline 
$5$ & $97\%$  & $97\%$ \\
\hline 
$6$ &$92\%$  & $91\%$ \\
\hline
\end{tabular}
\caption{Training and test accuracy results for identification of entanglement classes of four-qubit stabilizer states.}
\label{tab:stabilizer_states_reults}
\end{table}

\subsection{Identifying the $LU$-orbits of four-qubit graph states}\label{sec:lu-four-qubit}
We further trained the hybrid VQC to identify the LU orbits of four-qubit graph states. Since, LC group is a finite subset of the LU group, learning the LU orbits of quantum states is a much bigger task. Furthermore, the set of all states that can be reached by applying LU operations to graph states is continuous, and therefore this LU orbit contains an uncountable number of states. This identification of LU orbits is thus a big step up from the identification of LC orbits. 

Similar to the LC case, first, all graph states were generated, followed by the application of randomly generated LU operators to create sample states from their LU orbit. The \texttt{unitary\_group} statistical function from SciPy was used to generate Haar distributed random LU operators for this purpose. Once again, half of the samples were taken from the orbit to be identified and the rest from the other five orbits. The training and test accuracies for this case are shown in TABLE \ref{tab:lu_states_reults}.

\begin{table}[!ht]
\begin{tabular}{|c|c|c|c|}
\hline
Class N° & Training Accuracy  & Test Accuracy \\
\hline
$1$ & $94\%$ & $93\%$\\
\hline 
$2$   & $95\%$          & $94\%$ \\
\hline 
$3$ &$90\%$ & $89\%$ \\
\hline 
$4$ & $95\%$ & $95\%$ \\
\hline 
$5$ & $90\%$  & $89\%$ \\
\hline 
$6$ &$90\%$  & $90\%$ \\
\hline
\end{tabular}
\caption{Training and test accuracy results for identification of LU orbits of four-qubit graph states.}
\label{tab:lu_states_reults}
\end{table}

It should be noted that until now the samples have been restricted to states within the LU equivalence classes of four-qubit graph states, and therefore the VQC is largely ignorant of states that do not fall into any of the six classes. In light of that, we also trained our classifier to recognize the LU orbits of four-qubit graph states from the entire four-qubit Hilbert space. For this purpose, $50\%$ of the samples were taken from the LU equivalence class to be identified and the other $50\%$ were random four-qubit states generated by choosing random complex values for all $16$ elements of the amplitude vector. The training and test accuracies for this type of classification are shown in TABLE \ref{tab:lu_states_reults_2}.

\begin{table}[!ht]
\begin{tabular}{|c|c|c|c|}
\hline
Class N° & Training Accuracy  & Test Accuracy \\
\hline
$1$ & $95\%$ & $94\%$\\
\hline 
$2$   & $93\%$          & $91\%$ \\
\hline 
$3$ &$100\%$ & $99\%$ \\
\hline 
$4$ & $89\%$ & $88\%$ \\
\hline 
$5$ & $98\%$  & $97\%$ \\
\hline 
$6$ &$100\%$  & $99\%$ \\
\hline
\end{tabular}
\caption{Training and test accuracy results for identification of LU orbits of four-qubit graph states against the full four-qubit Hilbert space.}
\label{tab:lu_states_reults_2}
\end{table}

The results for this type of classification show better accuracy compared to those in TABLE \ref{tab:lu_states_reults} for most of the classes. At first glance, this may appear counterintuitive, since Table \ref{tab:lu_states_reults} presents a comparison between a single orbit and the other five four-qubit graph state LU-orbits, whereas Table \ref{tab:lu_states_reults_2} compares one orbit with all remaining states. However, when a state is randomly selected from the four-qubit Hilbert space, it is typically a generic state whose entanglement pattern is most likely SLOCC-equivalent to the generic $G_{abcd}$ four-qubit state, according to the classification of \cite{verstraete}. Therefore, in Table \ref{tab:lu_states_reults_2}, the classifier is trained to distinguish between two types of entanglement, the chosen class and the generic one, whereas in Table \ref{tab:lu_states_reults}, the dataset encompasses a wider variety of entanglement types. Consequently, the classifier performs significantly better when discriminating between two distinct types of entanglement than when attempting to identify a specific type within a diverse set of entanglement structures.

\subsection{Scaling to Larger Systems}
\label{sec:star_graph}
\begin{figure}[h!]
  \centering
  \subfigure[]{
    \includegraphics[width=0.13\textwidth]{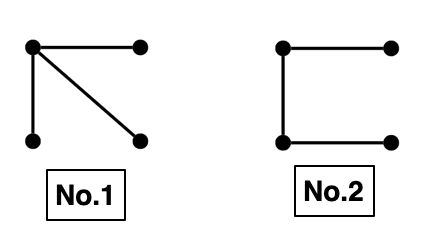}
    \label{fig:LU_4_qubit_graphs}
  }
  \subfigure[]{
    \includegraphics[width=0.32\textwidth]{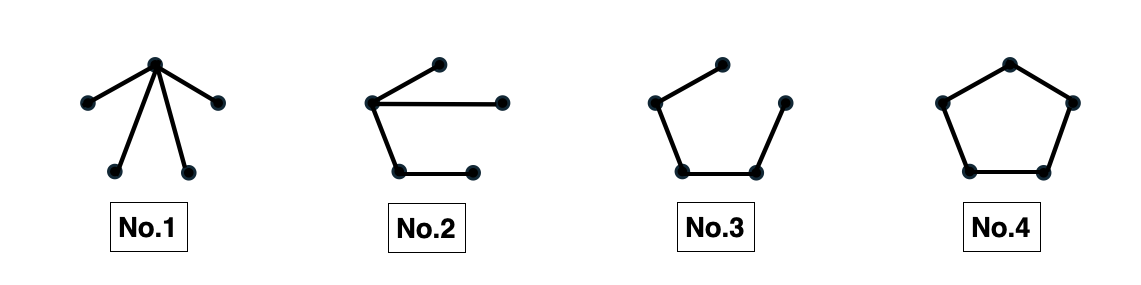}
    \label{fig:LU_5_qubit_graphs}
  }
  \subfigure[]{
    \includegraphics[width=0.43\textwidth]{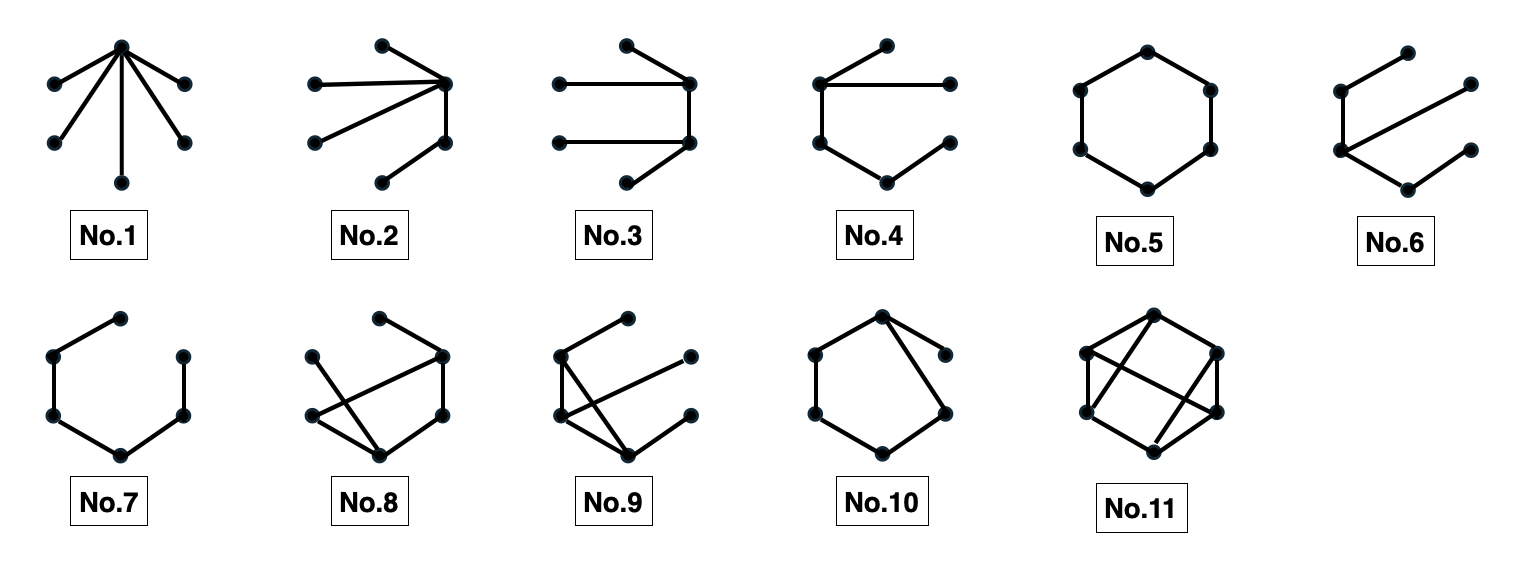}
    \label{fig:LU_6_qubit_graphs}
  }
  
  \caption{List of connected graphs that are not equivalent under LU transformation and graph isomorphism for 4 (a), 5 (b), and 6 (c) qubits \cite{hein2}. The star graph state for each system is the state titled No.1 in the corresponding subfigure.}
  \label{fig:LU_connected_graphs}
\end{figure}

In order to observe how this entanglement classification using hybrid VQC might scale up to systems with larger number of qubits, the LU orbit of the star graph state for 4, 5, and 6 qubit systems was classified against the rest of their LU orbits with the classification limited only to connected graphs. Connected graphs are graphs with a path of distinct edges connecting every pair of vertices \cite{hein2}. A connected graph state corresponds to genuine multi-partite entanglement and any graph state that is not connected can be given by the product of connected ones \cite{claudet}. Therefore, it is often sufficient to focus on connected graphs for entanglement classification. FIG. \ref{fig:LU_connected_graphs} shows all the connected graphs for 4, 5, and 6 qubits that are not equivalent under graph isomorphism and LU transformation.

The dataset was generated as described in Section \ref{sec:lu-four-qubit}. First, all the connected graphs shown in FIG. \ref{fig:LU_connected_graphs} were generated, followed by the application of randomly generated LU operators to each graph, thus creating sample states from their respective LU equivalence classes. Again, the \texttt{unitary\_group} function from SciPy was used to generate Haar distributed LU operators. For each system, half of the samples were picked from the LU orbit of its star graph state and the other half were picked uniformly from the LU orbits of the rest of its connected graphs shown in FIG. \ref{fig:LU_connected_graphs}.

The results of this classification are shown in Table \ref{tab:star_graph_state_results}. It can be observed that, as the size of the system increases from 4 to 5 to 6 qubits, the final accuracy values stay mostly consistent and drop only by a few percentage points. However, the number of neurons in the hidden layers of the classical NN as well as the number of samples needed to achieve these accuracy values increase noticeably with the size of the system. This can be attributed not just to the fact that the entanglement correlations become more complicated but also to the increase in the number of LU equivalence classes. As the system size increases from 4 to 5 to 6 qubits, the number of equivalence classes doubles and then rises to 11 \cite{hein2}. This number increases to 26, 101, 440, 3132, and 40457 classes for 7, 8, 9, 10, and 11 qubits respectively \cite{cabello}.

We terminated our classification results at 6 qubits due to limited computational resources. However, the results shown in Table \ref{tab:star_graph_state_results} indicate that our classification method would scale up well to larger systems, provided the computational resources are available.
\begin{table}[!ht]
\begin{tabular}{|c|c|c|c|c|}

\hline
System & Training  & Test  & No. of  & Classical NN \\
Size & Accuracy  & Accuracy & Samples & Architecture \\
\hline
$4$ & $99\%$ & $99\%$ & $9600$ & $(200,50)$\\
\hline 
$5$   & $97\%$          & $97\%$& $14400$ & $(300,50)$\\
\hline 
$6$ &$96\%$ & $96\%$ & $32400$ & $(500,50)$\\
\hline 
\end{tabular}

\caption{Training and test accuracy results for the classification of 4, 5, and 6 qubit star graph state against the rest of connected LU equivalence classes. The second last column shows the number of training samples used. Note that the number of training and test samples was always kept equal. The last column shows the number of neurons in the hidden layers of the classical post-processing NN with each layer separated by a comma. The rest of the hyperparameters and dataset features not mentioned here were kept consistent across all three cases.}
\label{tab:star_graph_state_results}
\end{table}

\section{Conclusion}
\label{conclusion}

In this paper we have presented a hybrid variational quantum circuit approach to entanglement classification. Our work shows that a hybrid VQC with post-processing through a classical NN does an excellent  job of learning highly non-linear entanglement orbits. We demonstrated this for the case of four-qubit graph states by first showing that a hybrid VQC is able to correctly classify all possible four-qubit graph states into their six distinct equivalence classes with $98\%$ accuracy or above. Following that, we show that this classifier is capable of recognizing and classifying all states in the LC orbit of these four-qubit graph states (the full finite set of stabilizer states) with $90\%$ accuracy or higher. The classifier also shows excellent performance for the even more complicated task of learning the LU orbits of these four-qubit graph states. We show that the hybrid VQC can learn to identify states from each of the six equivalence classes with accuracy $\geq88\%$. Finally, we also show that our approach scales up well to larger systems, showing strong numerical evidence for systems with $\leq$ 6 qubits.

These results highlight the robustness of our approach, especially when scaled up to more complicated classification schemes. Furthermore, the advantages of this approach, which include the lack of a need for full information of the quantum system, polynomial scaling of resources with the number of subsystems, and ready execution on currently available NISQ era hardware, make it an extremely effective and practically useful method of entanglement classification when combined with the high accuracy of our results. A review of prior work on the use of variational quantum classification methods for entanglement classification also highlights the uniqueness of our approach since training a hybrid VQC to learn non-linear orbits for classification, to our knowledge, has not been performed yet.

Several avenues remain open for future research in this area. First, the scalability of our approach to larger systems beyond 6 qubits needs to be addressed. Furthermore, while our research focused on graph states, this approach could be extended to the more tractable and commonly used SLOCC classification of all pure states. Finally, the performance of this classifier on actual quantum hardware and the corresponding impacts of noise and decoherence on its performance need to be explored. Investigating these possibilities could provide new insights into the practical classification of entangled states and its myriad applications in quantum information.

\section*{Acknowledgement}
This work was supported by the Graduate school EIPHI (contract ANR-17-EURE-
0002) and the QuanTEEM Erasmus Master program. All codes as well as details of implementation are available at \url{https://github.com/Hamna-Aslam3/HybridVQC-stabilizer-state-classification}.

\end{document}